\documentclass[aps,pre,twocolumn,superscriptaddress]{revtex4}

\usepackage{graphicx}
\usepackage{subfigure}
\begin{document}

\title{Monitoring Diffusion of Reptating Polymer Chains by Direct Energy Transfer
Method: a Monte Carlo Simulation}

\author{Erkan T\"uzel\footnote[2]{Present address: University of Minnesota, School of Physics and Astronomy,
116 Church St. SE, Minneapolis, MN, 55455, USA}}
\affiliation{Department of Physics, Faculty of Sciences and Letters, \\
Istanbul Technical University, Maslak 80626, Istanbul, Turkey}
\affiliation{Department of Physics, Faculty of Sciences and Letters, \\
I{\c s}{\i }k University, Maslak, 80670, Istanbul, Turkey}
\author{ K. Batuhan K{{\i}}sac{{\i}}ko{\u g}lu}
\affiliation{Department of Physics, Faculty of Sciences and Letters, \\
Istanbul Technical University, Maslak 80626, Istanbul, Turkey}
\author{{\"O}nder Pekcan}
\affiliation{Department of Physics, Faculty of Sciences and Letters, \\
Istanbul Technical University, Maslak 80626, Istanbul, Turkey}

\newcommand{\be}{\begin{equation}}
\newcommand{\ee}{\end{equation}}
\newcommand{\bea}{\begin{eqnarray}}
\newcommand{\eea}{\end{eqnarray}}
\newcommand{\bc}{\begin{center}}
\newcommand{\ec}{\end{center}}

\pagenumbering{arabic}

\begin{abstract}

A kinetic Monte Carlo method was used to simulate the diffusion of
reptating polymer chains across the interface. A time-resolved
fluorescence technique conjunction with direct energy transfer
method was used to measure the extend of diffusion of dye labeled
reptating polymer chains. The diffusion of donor and acceptor
labeled polymer chains between adjacent compartments was randomly
generated. The fluorescence decay profiles of donor molecules were
simulated at several diffusion steps to produce mixing of the polymer
chains. Mixing ratios of donor and acceptor labeled polymer chains
in compartments were measured at various stages (snapshots) of diffusion.
It was observed that for a given molecular weight, the average
interpenetration contour length was found to be proportional to
the mixing ratio. Monte Carlo analysis showed that curvilinear
diffusion coefficient is inversely proportional to the weight of
polymer chains during diffusion.
\end{abstract}

\maketitle
\newpage
\section{Introduction}

The diffusion of polymer chains across polymer-polymer interfaces
has been of interest for more than a
decade.\textsuperscript{\cite{pekcan1,winnik1}} One of the reason
for the interest is that polymer diffusion across an interface is
found to be important in technological processes, such as
sintering of polymer powders, development of the strength of
polymer powder by compression molding and annealing and the
formation of latex films. Latex-film formation has
been considered in the literature for over $50$ years and
is frequently used in the modern
industry.\textsuperscript{\cite{mazur,keddie}} The process of
latex film formation has been divided into several stages. The
generally accepted mechanisms consist of: (i) evaporation which brings
the particles into some form of close packing; (ii) deformation of
particles which leads to a structure without voids, although with
the original particles still distinguishable; and (iii) diffusion of polymer
chains across particle-particle boundaries, yielding a continuous
film with mechanical integrity. Voyutskii proposed that physical
contact between latex particles would not produce a mechanically
strong continuous film if no external effect is
applied.\textsuperscript{\cite{voyutskii}} In other words, in order
to obtain a stable film it is necessary that the segments of
polymer chains diffuse from one particle to another by
forming a strong linkage between them.  He described the process
of diffusion with the word of ``autohesion''. For several decades,
after Voyutskii's paper there have been only speculations about
the process of diffusion during latex film formation.  The
developments in neutron scattering techniques have enabled the carrying 
out of experiments to study diffusion of polymer molecules
across particle-particle boundaries.  Important progress was
made when small angle neutron scattering (SANS) technique was
applied to study diffusion between particles of hydrogenated
and deuterated acrylic latex
particles.\textsuperscript{\cite{hahn1}} Later, it was reported
that increasing molecular weight and incompatibility lower the
diffusion rate.\textsuperscript{\cite{hahn2}} SANS was employed to
measure the extend of diffusion in polystyrene (PS) latex having
high molecular weight and small particle size, where an increase
in the radius of gyration ($R_g$) of the polymer was observed when
the system was heated above its
$T_g$.\textsuperscript{\cite{kim1}}

The non-radiative direct energy transfer (DET) method conjunction with
time resolved fluorescence (TRF) tecnique was first used to study
polymer diffusion across particle-particle
boundaries\textsuperscript{\cite{pekcan2}} to monitor
concentration profiles of donor and acceptor dyes attached to
polymers that are located initially in separate particles.  As
polymer diffusion occurs, mixing of donor (D) and acceptor (A)
dyes can be measured by an increase in energy transfer between
them. Early measurements using DET on poly (methyl methacrylate)
(PMMA) latex particles prepared by nonaqueous dispersion
polymerization found diffusion coefficients of the
order of $10^{-15} cm^2 s^{-1}$ at temperatures between $400-450$
K. The film formation of a poly (butyl methacrylate) latex
prepared via emulsion polymerization in water, was studied using the
same technique. The diffusion coefficients ($10^{-16} cm^2
s^{-1}$) were determined using spherical diffusion model, and
found to be dependent on both time and temperature. A decrease in
diffusion rate with time was attributed to the effect of low
molecular weight chains near the particle-particle surface
dominating at early times. The DET method of data analysis was
later developed for melt pressed PMMA
particles.\textsuperscript{\cite{wang1}} It is observed that mass
transfer increased with time to a power of $0.5$ as in Fickian
diffusion model for low molecular weight polymers. For high
molecular weights, there is a distinct $0.25$ power dependence
that cannot be explained by a Fickian or reptation model.

Two reviews\textsuperscript{\cite{wang2,pekcan1}} summarized
experimental approaches and results in studying diffusion in latex
systems up to the year of 1993 and 1994, respectively.  Both papers
reviewed the evidence of diffusion of polymers at
particle-particle junction using transmission electron microscopy, SANS, atomic force microscopy
and fluorescence techniques. Data analysis using DET and a fluorescence 
technique was later
improved by taking into account the donor and acceptor
concentration profiles during polymer diffusion, where a uniform
acceptor concentration was considered around a donor in thin
slices or shells.\textsuperscript{\cite{farinha1,oneil}} Further a
model for DET was developed which considered the heterogeneity in
the donor and acceptor concentration
profiles,\textsuperscript{\cite{farinha2}} where the diffusion
coefficient of polymer chains obtained by different DET models.
DET studies in latex blends, where one phase is far below its
$T_g$ and does not undergo any significant diffusion, found 
that the magnitude of energy transfer is proportional to the
interfacial area. It was observed that organic solvents, which
plasticize the latex, enhance the rates of polymer diffusion and
the diffusion rate increased in the presence of
cosurfactants.\textsuperscript{\cite{juhue}} At the times longer
than the tube renewal time (escape time), the activation energy
for diffusion was found 30 kcal/mol in the presence of
cosurfactants, which is 48 kcal/mol in the neat polystyrene. Monte
Carlo studies on diffusion of donor and acceptor dyes between the
adjacent compartments were modeled by using Brownian
motion.\textsuperscript{\cite{tuzel,gunturk1,gunturk2}}

In this work Kinetic Monte Carlo (KMC) method was used to simulate
the diffusion of polymer chains between the adjacent compartments
of a cube where free energies and the potentials between molecules
were not considered. These chains are labeled with either donor or
acceptor dyes. Reptation of polymer chains were simulated using
KMC method which produced more realistic diffusion results
compared to previous
studies.\textsuperscript{\cite{tuzel,gunturk1,gunturk2}} The
diffusion of donor and acceptor labeled polymer chains between
adjacent compartments was randomly generated where each chain
reptates according to de Gennes model.\textsuperscript{\cite{degennes}} The decay of the donor
fluorescence intensity, $I(t)$ by DET was simulated at several
diffusion steps and a gaussian noise was added to generate the
time resolved fluorescence decay data. $I(t)$ decays were then
fitted to the phenomenological equation

%%%%%%%%%%%%%%%%%%%%%%%%%    FIGURE 1  %%%%%%%%%%%%%%%%%%%%%%%%%%%%%%%%%%%%
\begin{figure}
\bc
\leavevmode
\includegraphics[width=8cm,angle=0]{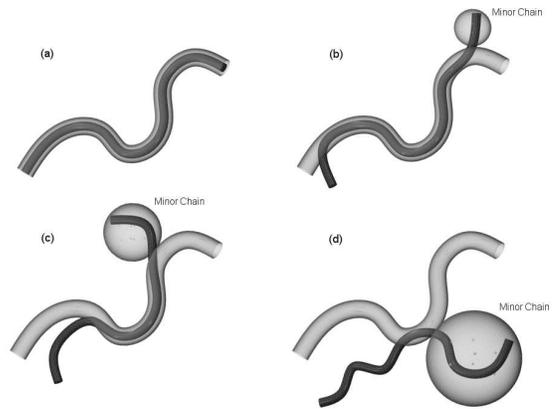}
\caption{\small{A cartoon representation of a reptating
polymer chain in a tube at different times (a,b,c and d). The
memory of the initial tube is lost in Figure 1d. The picture also
demonstrates the motion of minor chain during reptation.
}}
\ec
\end{figure}
%%%%%%%%%%%%%%%%%%%%%%%%%%%%%%%%%%%%%%%%%%%%%%%%%%%%%%%%%%%%%%%%%%%%%%%%%%%

\be
\frac{I(t)}{I(0)}=B_1\exp(-t/{\tau_0}-C(\frac{t}{\tau_0})^{1/2})+
B_2\exp(-t/{\tau_0}) \label{pnemon}
\ee

\noindent where $\tau_0$ is the fluorescence lifetime of donor and
$B_1$ and $B_2$ are the pre-exponential factors. Here it is useful
to define the mixing ratio $K$ representing the order of mixing
during diffusion of the donor and the acceptor labeled polymer
chains as

\be K=\frac{B_1}{B_1+B_2}  \label{k} \;\;.\ee

\noindent $K$ was measured in terms of $B_1$ and $B_2$ for various
molecular weights, $M$ of polymer chains. Results were interpreted
in terms of average interpenetration contour length, $\ell(t_r)$
and curvilinear diffusion coefficient, $D_1$.

\section{Theoretical Considerations}

\subsection{Molecular Motion of Polymer Chains}

The motion of individual chains in bulk polymeric materials or
concentrated solutions of linear random-coil (Gaussian) chains has
been modeled by the reptation theory of de
Gennes\textsuperscript{\cite{degennes}} and
Edwards.\textsuperscript{\cite{edwards}} In this model the polymer
chain is confined to a tube which presents topological constraints
to lateral motion of monomers imposed by the neighboring chains
via entanglements. The motion of the chain is restricted to the
curvilinear length of the tube, which is shown in Figure 1. As
shown in Figure 1a, at time zero, the chain is in its initial tube
which performs Brownian back and forth motion. Since the chain
ends are free to move in any direction away from the tube, the
memory of the initial tube position in space is gradually lost as
shown in Figures 1b and 1c. Figure 1d presents the final stage of
the reptation, i.e. just before the total escape of the chain from
its original tube. At the time $T_r$, the escape time, the chain
escapes or forgets its original configuration. As the chain is in
a bulk polymeric system, it always creates a new tube after escaping
from the previous one. The length of chain, $\ell(t_r)$ which
escapes from the initial tube is also a random-coil chain which
obeys Gaussian statistics and called the minor
chain.\textsuperscript{\cite{kim2,wool}} During the process of
reptation the length of the minor chain grows which is represented
in Figure 1 by the growing spherical envelope. The mean square
escape length of the minor chains, $<\ell^2>$ is calculated by the
following relation\textsuperscript{\cite{kim2}}

%%%%%%%%%%%%%%%%%%%%%%%%%    FIGURE 2  %%%%%%%%%%%%%%%%%%%%%%%%%%%%%%%%%%%%
\begin{figure}
\bc
\leavevmode
\includegraphics[width=6cm,angle=0]{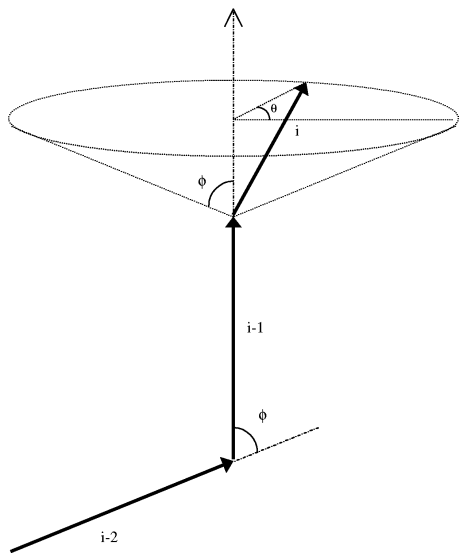}
\caption{\small{Schematic picture of the bond and
dihedral angles for the polymer chains. The bond angle $\phi$ is
equal to $60^o$, and the dihedral angles, namely $\theta$, can be
randomly chosen.
}}
\ec
\end{figure}
%%%%%%%%%%%%%%%%%%%%%%%%%%%%%%%%%%%%%%%%%%%%%%%%%%%%%%%%%%%%%%%%%%%%%%%%%%%

\be
<\ell^2> = 16 \pi^{-1}D_1 t_r \label{lsquare} \;\;.
\ee

\noindent where $t_r$ is the diffusion time and $D_1$ is the
curvilinear one-dimensional diffusion coefficient. The molecular
weight dependence of $D_1$ and $T_r$ are given by the following
relations

\be
D_1 \sim M^{-1} \label{d1_prop_invM}
\ee

\noindent
and

\be
T_r \sim M^{3} \label{tr_prop_m3}
\ee

\noindent
respectively.

The average interpenetration contour length, $\ell(t_r)$ of chain
segments which have diffused across the interface is obtained from
the minor chain model as\textsuperscript{\cite{wool}}

\be \ell(t_r) \sim M^{-1/2}t_r^{1/2} \label{interpenet}\ee

\be \ell_{\infty} \sim M \ee

\noindent This property has the same scaling relation as the minor
chain length, $<\ell>$.

\subsection{Direct Energy Transfer Method}

Time resolved fluorescence (TRF) in conjunction with DET method
monitors the extent of diffusion of donor (D) and the acceptor (A)
labeled polymer molecules. If the sample is made of a mixture of D
and A labeled polymer chains where the diffusion takes place,
after a period of time when the donor fluorescence profiles are
measured, each decay trace provides a snapshot of the extent of
diffusion.\textsuperscript{\cite{pekcan2}} This sample is
considered to be composed of three regions; unmixed D, unmixed A
and the mixed D - A region. The above model was first empirically
introduced by the two component donor fluorescence
decay.\textsuperscript{\cite{winnik2}}

%%%%%%%%%%%%%%%%%%%%%%%%%    FIGURE 3ab  %%%%%%%%%%%%%%%%%%%%%%%%%%%%%%%%%%%%
\begin{figure}
\bc
\leavevmode
\includegraphics[width=8cm,angle=0]{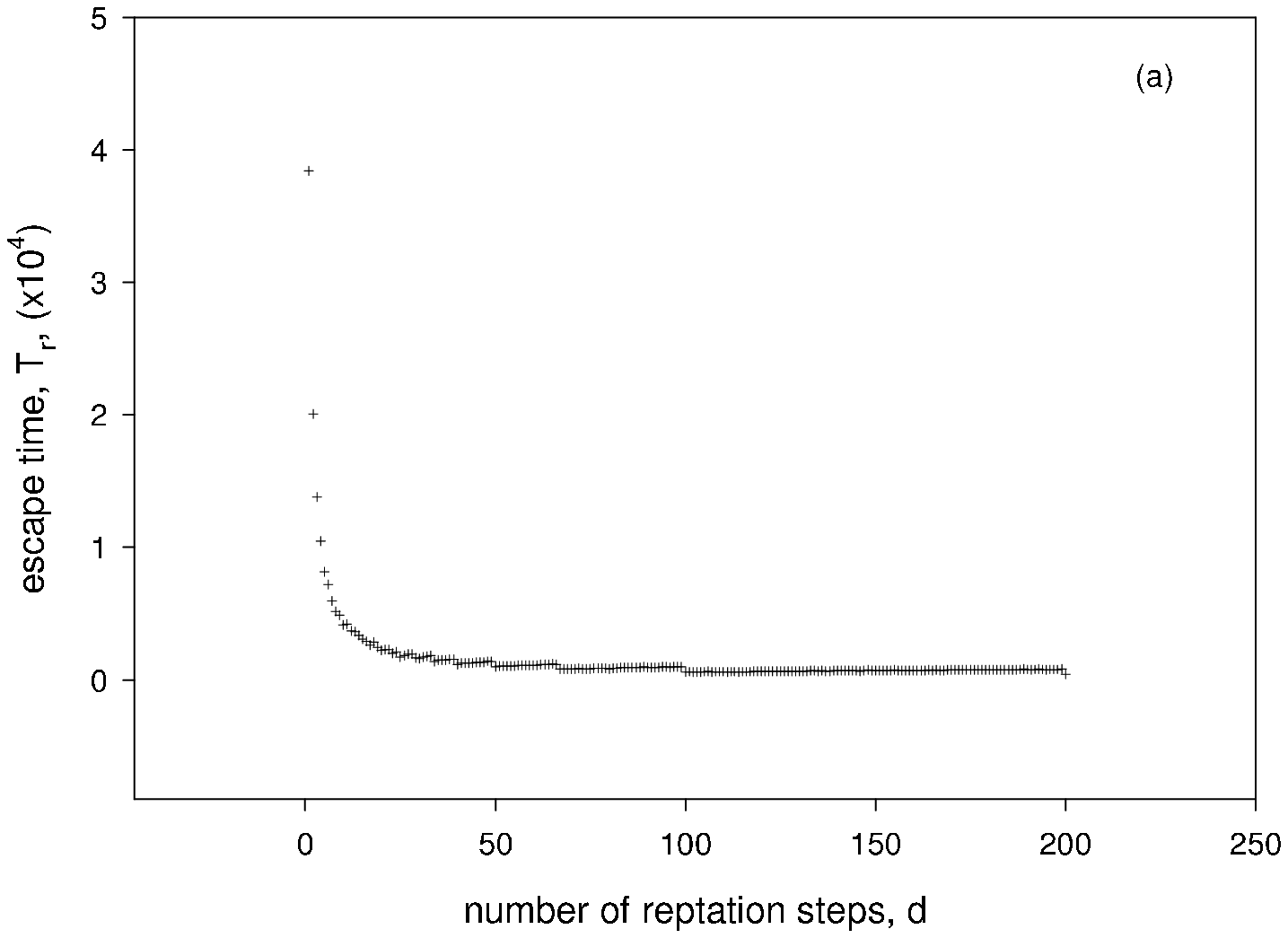}
\bigskip
\includegraphics[width=8cm,angle=0]{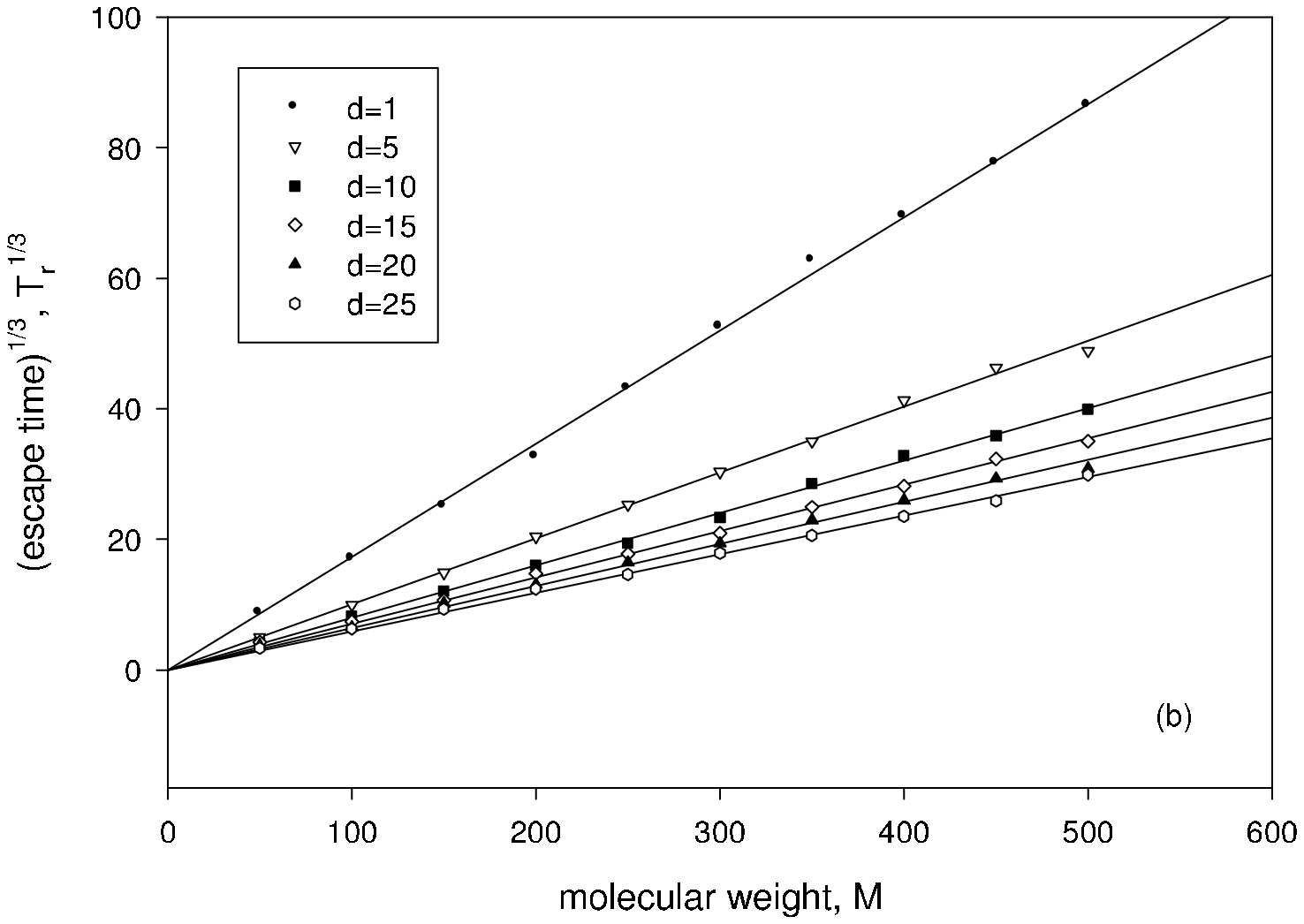}
\caption{\small{The plot of  a) escape time, $T_r$
against the number of reptation steps, $d$, b) $(T_r)^{1/3}$ versus molecular weight, $M$, for
different reptation steps, $d$, for a polymer chain containing 200 monomer units.
}}
\ec
\end{figure}
%%%%%%%%%%%%%%%%%%%%%%%%%%%%%%%%%%%%%%%%%%%%%%%%%%%%%%%%%%%%%%%%%%%%%%%%%%%%

%%%%%%%%%%%%%%%%%%%%%%%%%    FIGURE 4ab  %%%%%%%%%%%%%%%%%%%%%%%%%%%%%%%%%%%%
\begin{figure}
\bc
\leavevmode
\includegraphics[width=8cm,angle=0]{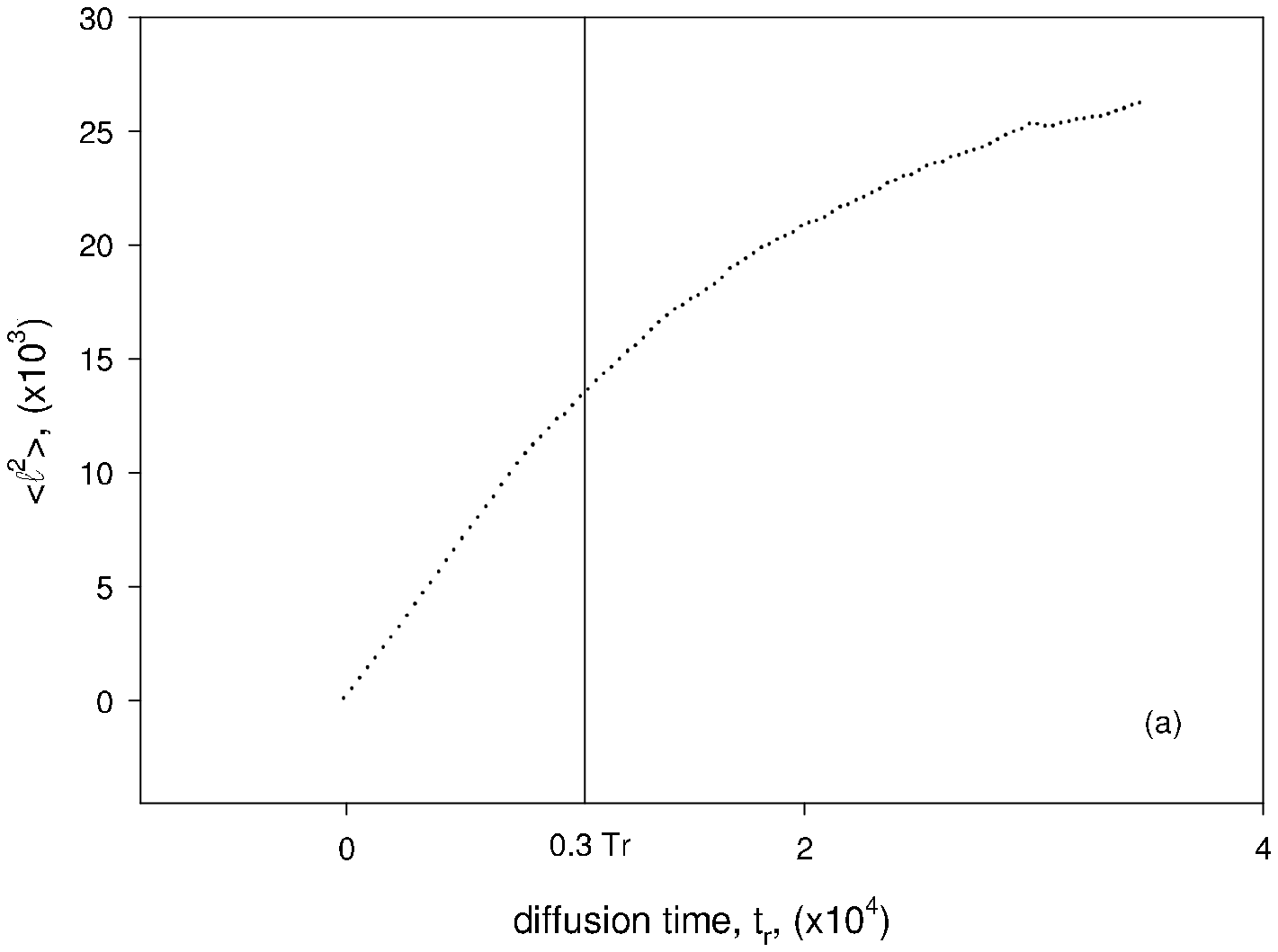}
\bigskip
\includegraphics[width=8cm,angle=0]{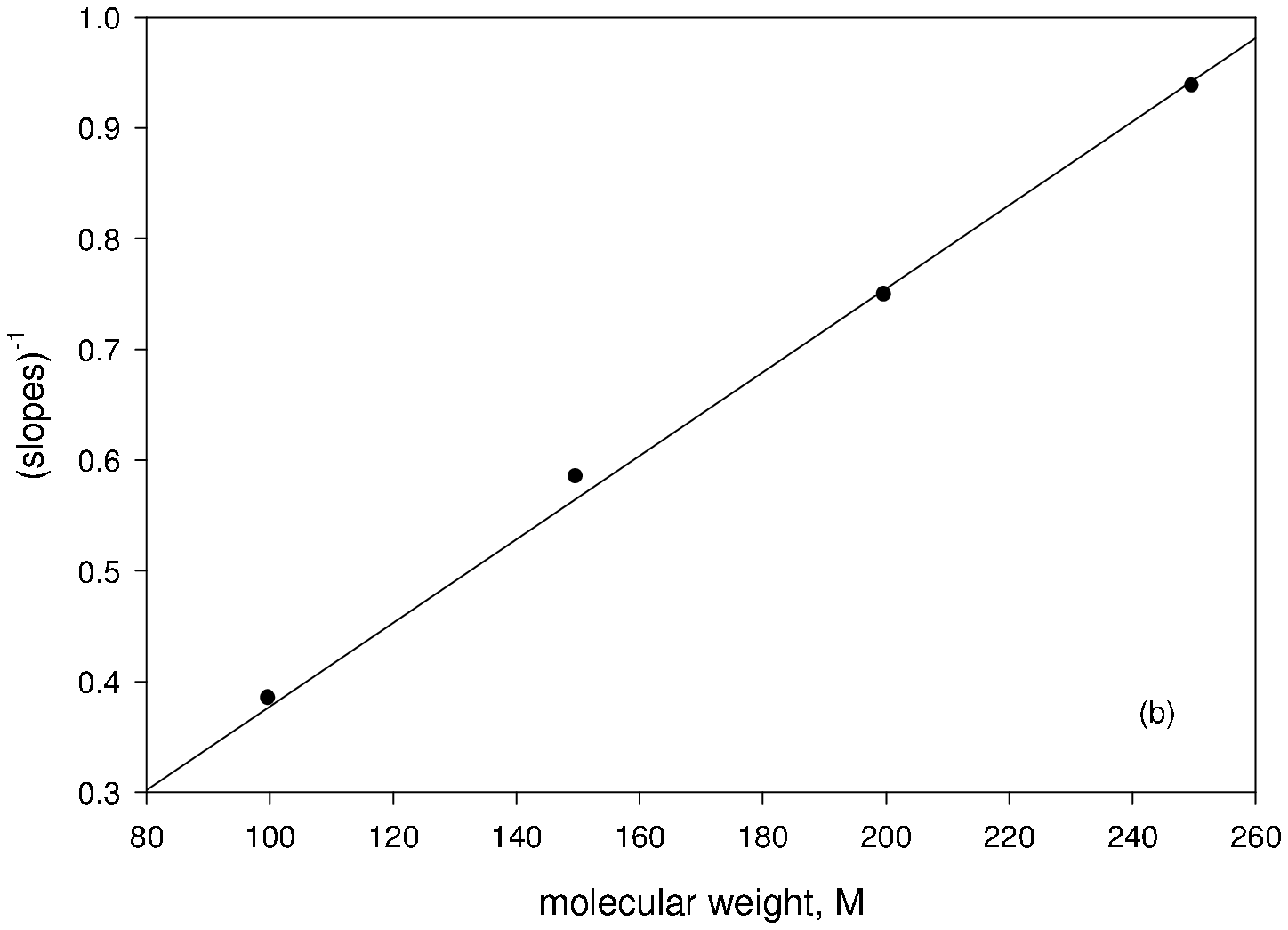}
\caption{\small{The plot of a) mean square escape
length, $<{\ell}^2>$ of minor chain versus diffusion time, $t_r$, b) The inverse slope of $<{\ell}^2>$ versus
molecular weight, $M$. For a polymer chain of 200 monomer units.
}}
\ec
\end{figure}
%%%%%%%%%%%%%%%%%%%%%%%%%%%%%%%%%%%%%%%%%%%%%%%%%%%%%%%%%%%%%%%%%%%%%%%%%%%%

When donor dyes are excited using a very narrow pulse of light,
the excited donor returns to the ground state either by emitting a
fluorescence photon or through the nonradiative mechanism. For a
well behaved system, after exposing the donors to a short pulse
of light, the fluorescence intensity decays exponentially with
time. However, if acceptors are present in the vicinity of the
excited donor, then there is a possibility of DET from the excited
donor to the ground state acceptors. In the classical problem of
DET, neglecting back transfer, the probability of the decay of the
donor at $r_k$ due to the presence of an acceptor at $r_i$ is
given by\textsuperscript{\cite{forster}}

\be
P_k(t) = exp[-t/{\tau_0}-{w_{ik}}t] \label{pk}
\ee
	
\noindent where $w_{ik}$ is the rate of energy transfer given by
F{\"o}rster as

\be
w_{ik} = {\frac {3}{2}}{{\kappa}^2}{\frac{1}{{\tau}_0}}{({\frac{R_0}{r_{ik}}})}^6 \;\; . \label{wik}
\ee

\noindent Here $R_0$ represents the critical F{\"o}rster distance
and $\kappa$ is a dimensionless parameter related to the geometry
of interacting dipole. If the system contains $N_D$ donors and
$N_A$ acceptors, then the donor fluorescence intensity decay can
be derived from the equation (\ref{wik}) and given
by\textsuperscript{\cite{wang1,wang2}}

\bea
\frac{I(t)}{I(0)}&=&\exp(-t/{\tau_0})\frac{1}{N_D}\int{{n_D} (r_k)dr_k} \nonumber \\
&&\times \prod_{i=1}^{N_A} {\frac{1}{N_A}}\int{{n_A}{(r_i)}}dr_i\exp{(-w_{ik}t)} \;\; . \label{I1}
\eea

\noindent Here $n_D$ and $n_A$ represent the distribution
functions of donors and acceptors. In the thermodynamic limit
equation (\ref{I1}) becomes

\bea
\frac{I(t)}{I(0)}&=&\exp(-t/{\tau_0})\frac{1}{N_D}\int{{n_D}(r_k)dr_k} \nonumber \\
&\times& \exp({-\int{n_A(r_i)dr_i(1-\exp{(-{w_{ik}}t))}}})  \;\; . \label{I2}
\eea

%%%%%%%%%%%%%%%%%%%%%%%%%    FIGURE 5abc  %%%%%%%%%%%%%%%%%%%%%%%%%%%%%%%%%%%%
\begin{figure}
\bc
\leavevmode
\includegraphics[width=6.5cm,angle=270]{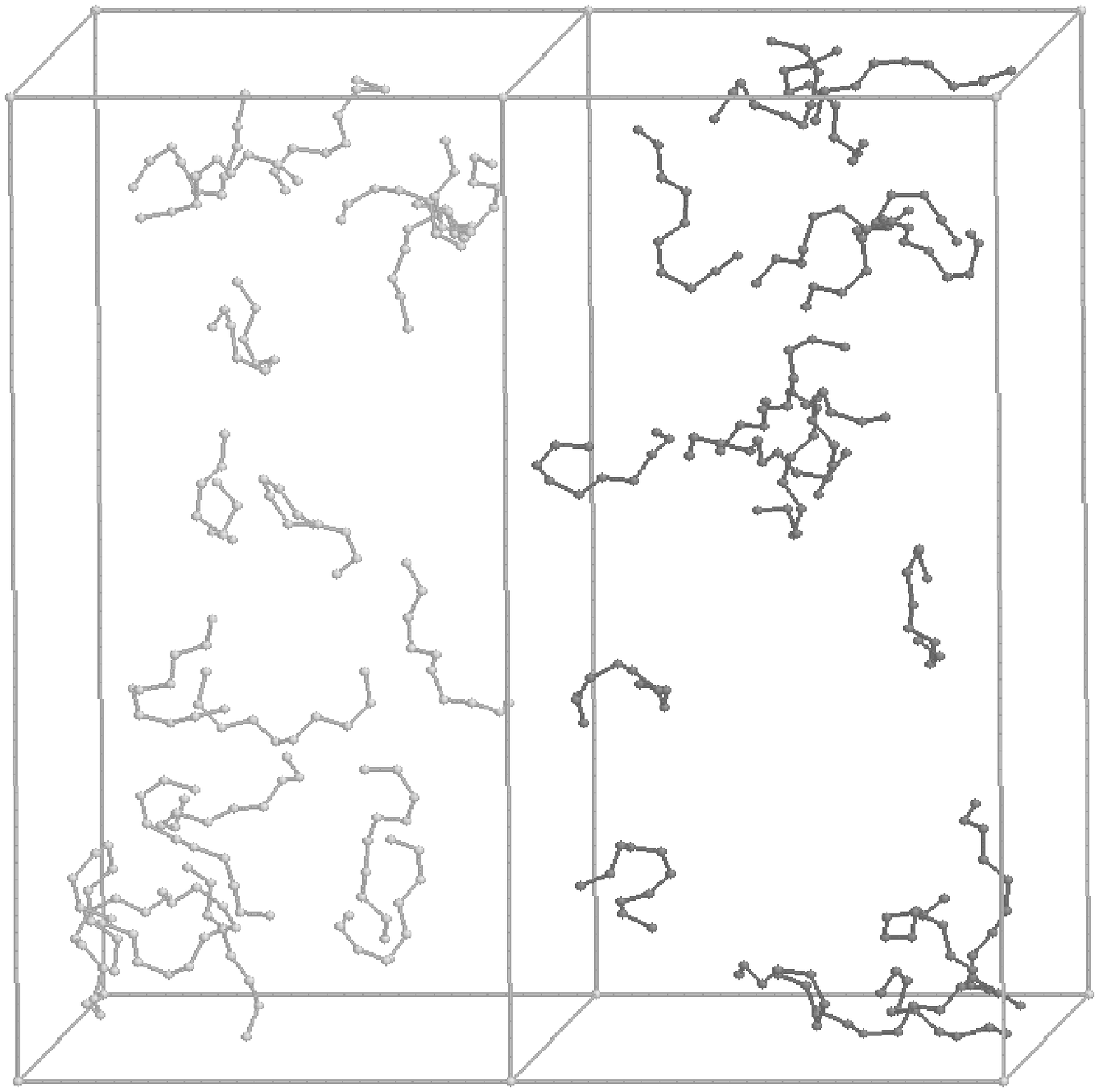}
\bigskip
\includegraphics[width=6.5cm,angle=270]{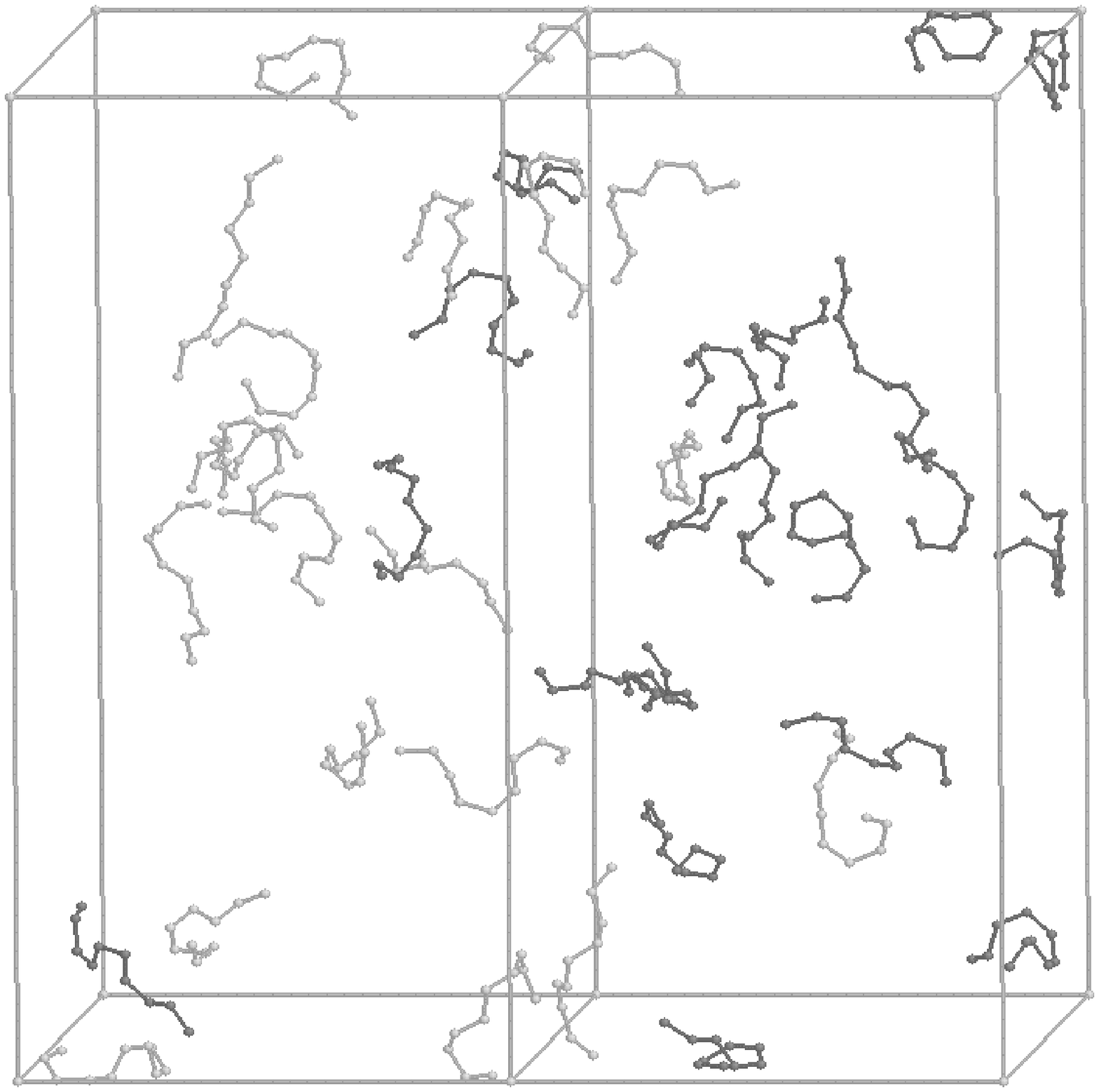}
\bigskip
\includegraphics[width=6.5cm,angle=270]{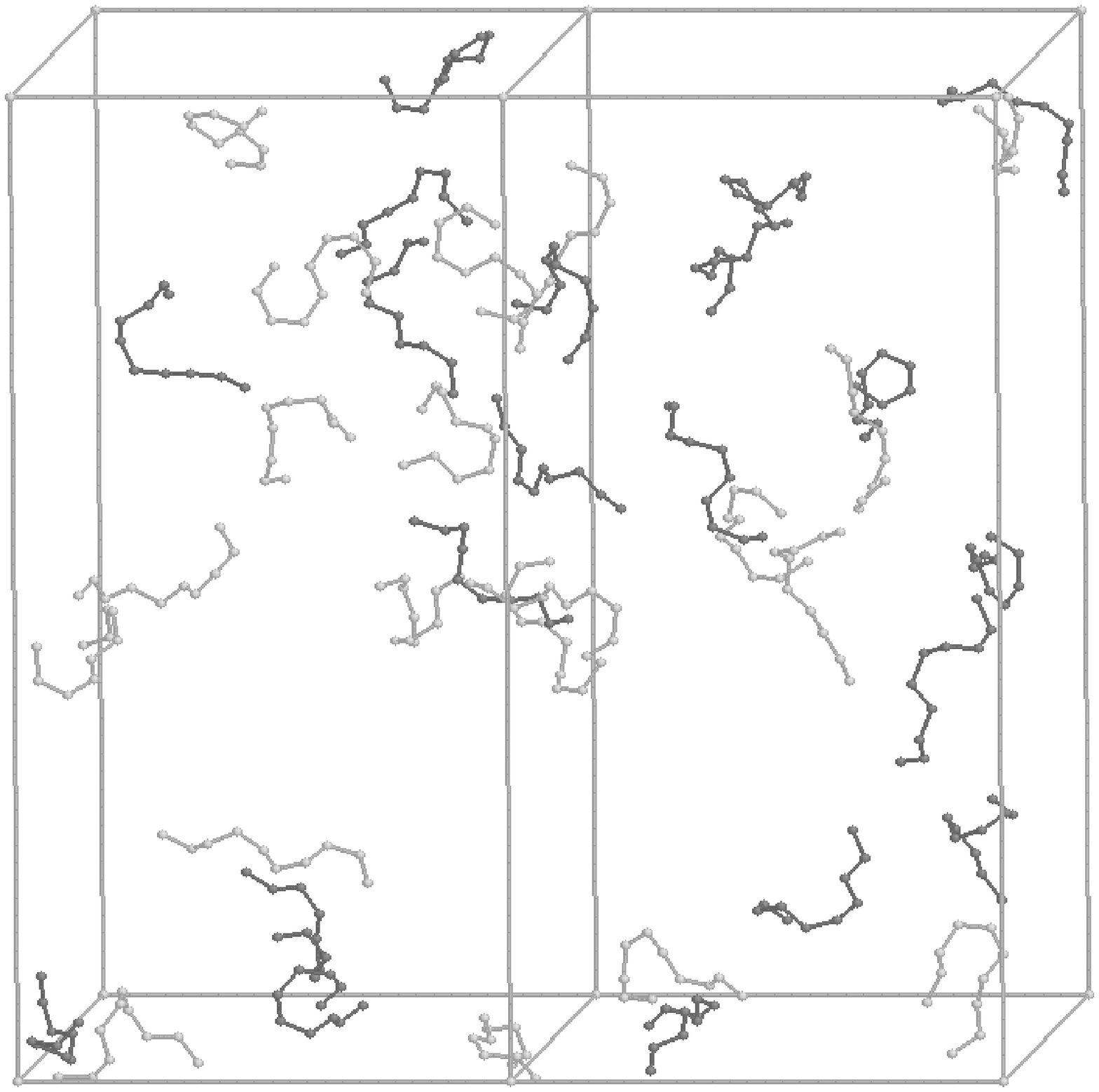}
\caption{\small{Snapshots of diffusion between adjacent
compartments of the cube with the size $L\times L\times L (L=30)$.
Donor, D labeled chains in the right compartment are presented in
black for clarity. The chains consist of $10$ units and the number
of chains in each compartment is taken to be $20$. The grey
colored chains represent the ones that are labeled by acceptors,
A.
}}
\ec
\end{figure}
%%%%%%%%%%%%%%%%%%%%%%%%%%%%%%%%%%%%%%%%%%%%%%%%%%%%%%%%%%%%%%%%%%%%%%%%%%%%

%%%%%%%%%%%%%%%%%%%%%%%%%    FIGURE 6ab  %%%%%%%%%%%%%%%%%%%%%%%%%%%%%%%%%%%%
\begin{figure}
\bc
\leavevmode
\includegraphics[width=8cm,angle=0]{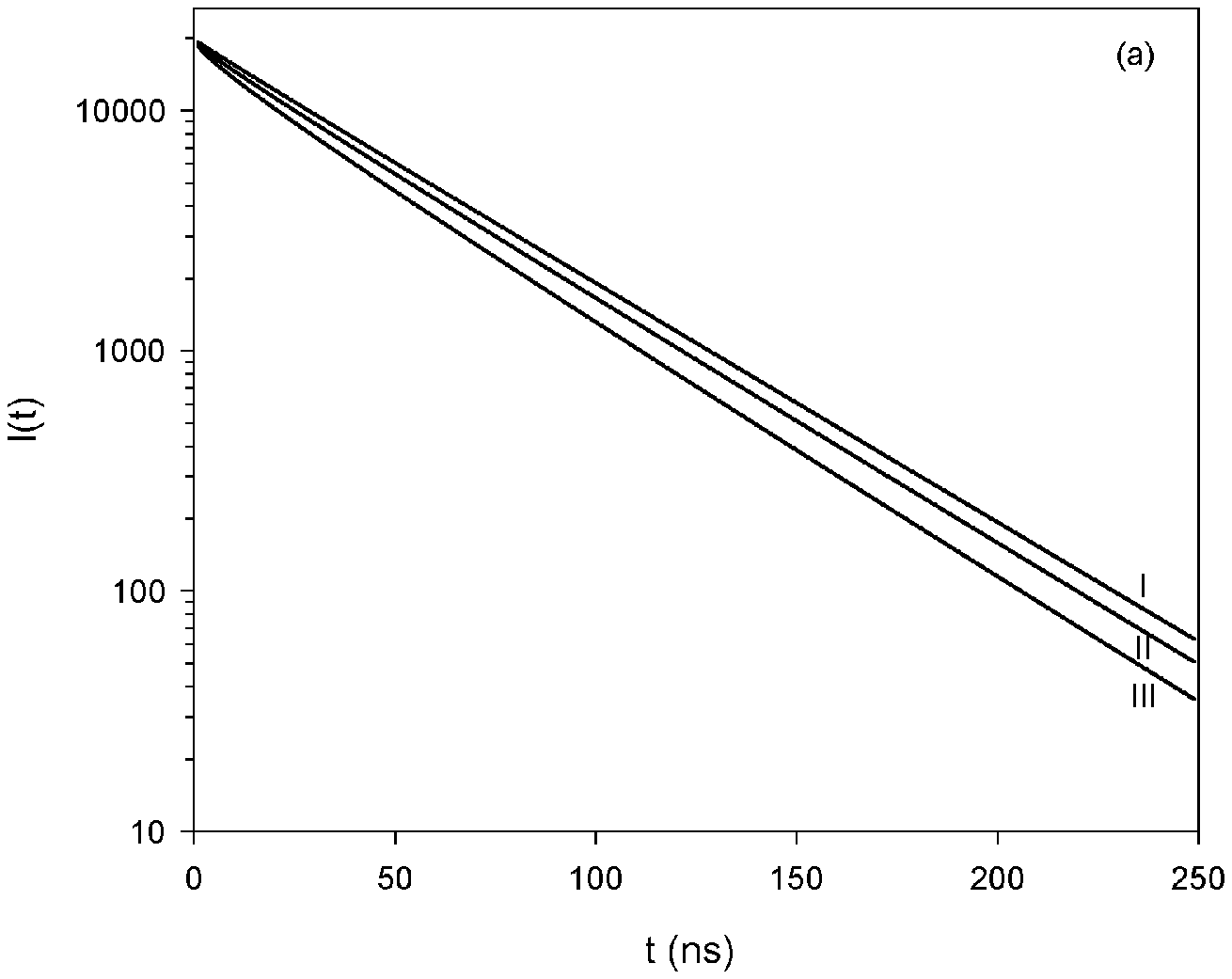}
\bigskip
\includegraphics[width=8cm,angle=0]{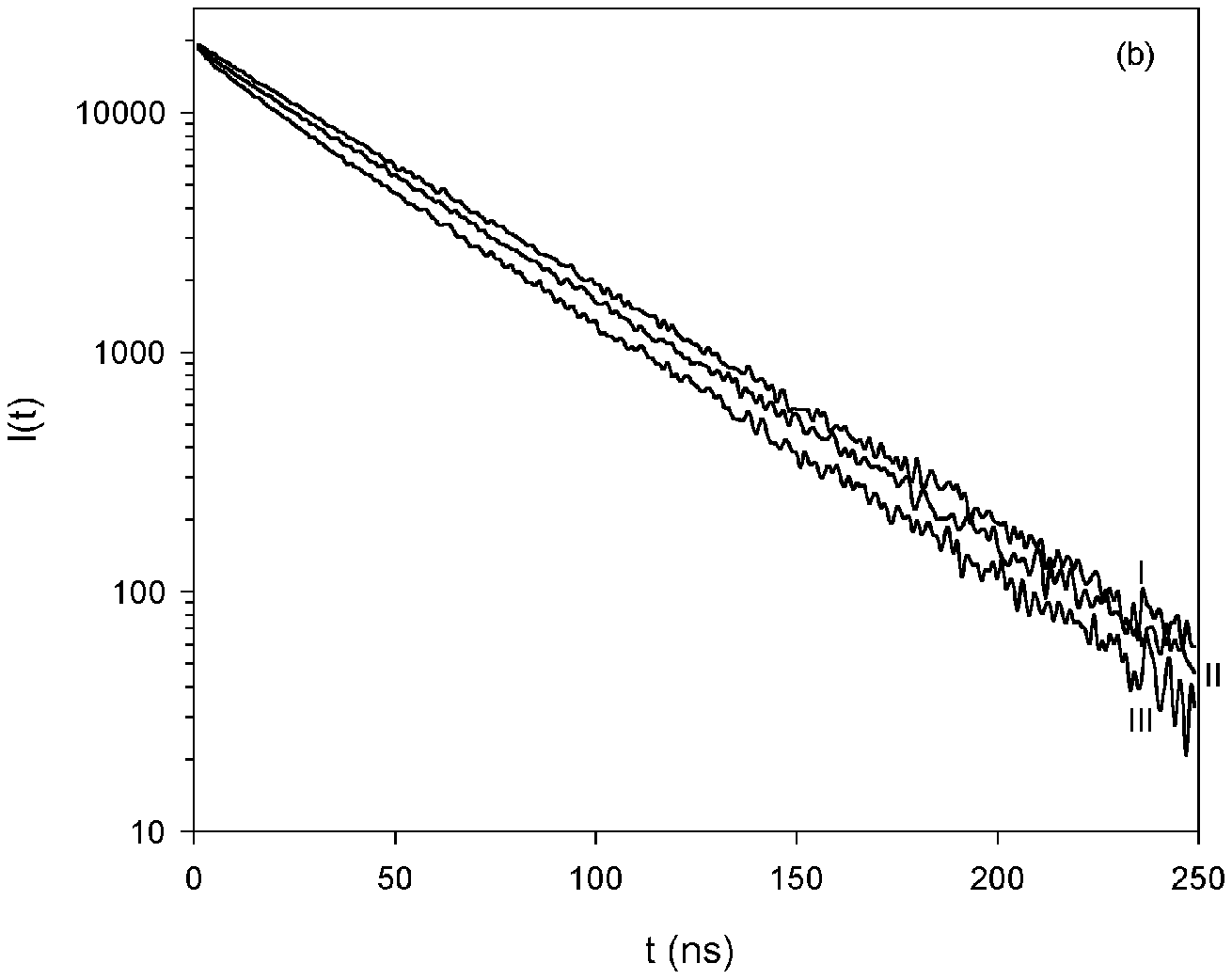}
\caption{\small{a) Donor decay profiles at several
diffusion steps for $M=100$ and $d=10$. Where the mixing ratios
are taken as I) $K=0.1$, II) $K=0.5$, III) $K=0.9$, b) Noisy decay profiles for the above picture.
}}
\ec
\end{figure}
%%%%%%%%%%%%%%%%%%%%%%%%%%%%%%%%%%%%%%%%%%%%%%%%%%%%%%%%%%%%%%%%%%%%%%%%%%%%

\noindent This equation can be used to generate donor decay
profiles by Monte-Carlo techniques. It is shown that the equation
(\ref{I2}) reduces to a more simple form which can be compared to
the experimental data.\textsuperscript{\cite{sperry}} The simplification
is summarized below for clarity. Changing to the
coordinate $r_{ik}=r_i-r_k$ leads to,

\bea
&&\frac{I(t)}{I(0)}=\exp(-t/{\tau_0})\frac{1}{N_D}\int{{n_D}(r_k)dr_k} \nonumber \\
&&\;\;\;\times \prod_{i=1}^{N_A} \int_{r_K}^{R_g-r_K}{{n_A}{({r_{ik}}+r_k)}}dr_{ik}
\exp{(-w_{ik}t)} \label{I3}
\eea

\noindent where $R_g$ is an arbitrary upper limit. Placing a
particular donor at the origin and assuming that the mixed and
unmixed regions are created during diffusion of D and A, equation
(\ref{I3}) becomes

\bea
\frac{I(t)}{I(0)}&=&B_1\exp(-t/{\tau_0})\prod_{i=1}^{N_A}\frac{1}{N_A}
\int_{0}^{R_g}{{n_A}(r_{ik})dr_{ik}} \nonumber \\
&&\times \exp{({-w_{ik}}t)}+B_2
\exp(-t/{\tau_0}) \label{I4}
\eea

\noindent
where

\be
B_{1,2}=\frac{1}{N_D}\int_{1,2}{n_D(r_k)dr(k)}
\ee

\noindent represent the fraction of donors in mixed and unmixed
regions respectively. The integral in equation (\ref{I4}) produces
a F{\"o}rster type of
function\textsuperscript{\cite{klafter,bauman}}

\be
\prod_{i=1}^{N_A} \frac{1}{N_A} \int_0^{R_g} n_A(r_{ik})dr_{ik}exp(-w_{ik}t)=
exp(-C({\frac{t}{\tau_0}})^{1/2})
\ee

\noindent where $C$ is proportional to acceptor concentration.
Eventually, one gets  equation (\ref{pnemon}) for the fluorescence
intensity.

\section{Results and Discussion}
%%%%%%%%%%%%%%%%%%%%%%%%%    FIGURE 7ab  %%%%%%%%%%%%%%%%%%%%%%%%%%%%%%%%%%%%
\begin{figure}
\bc
\leavevmode
\includegraphics[width=8cm,angle=0]{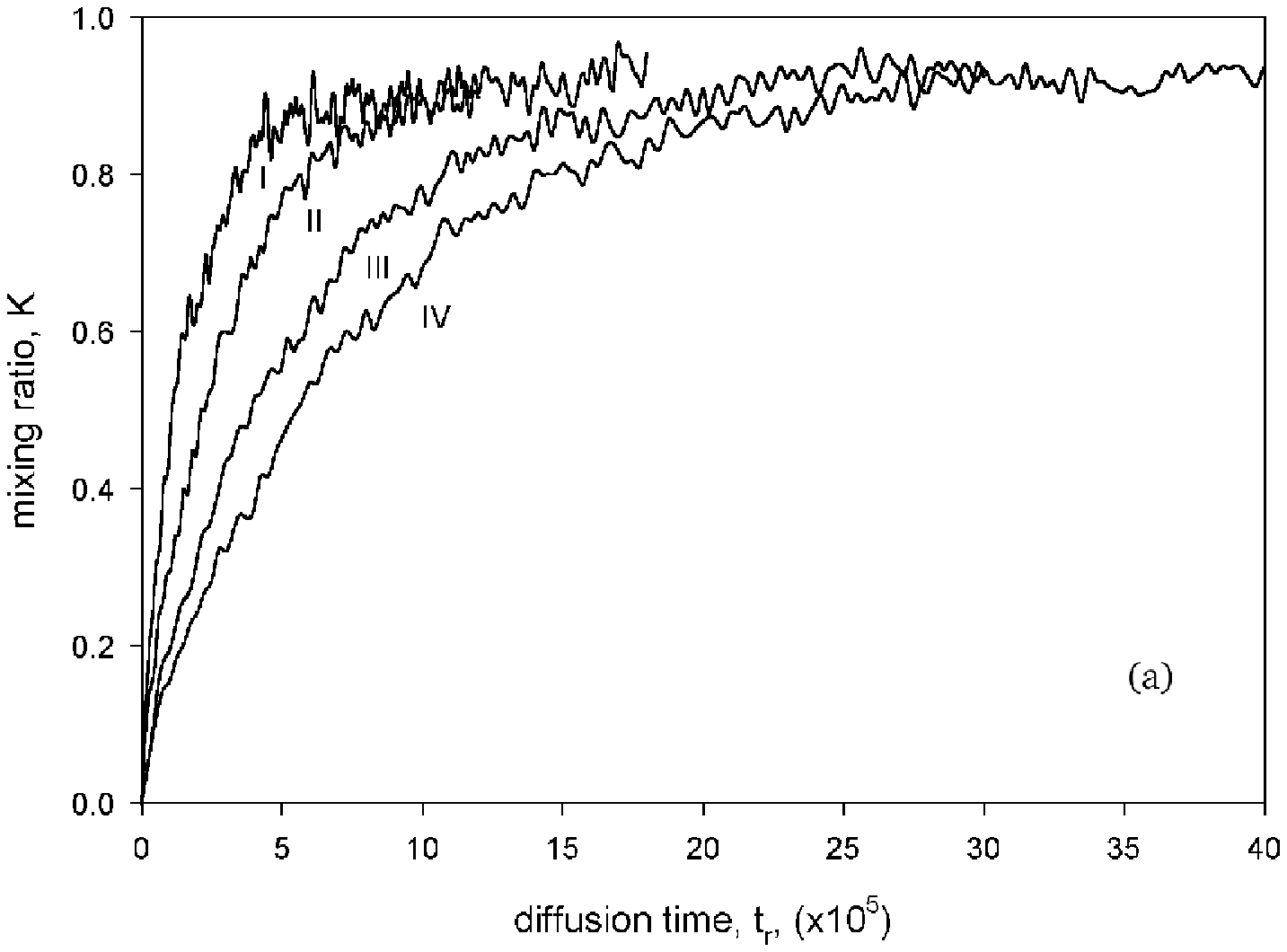}
\bigskip
\includegraphics[width=8cm,angle=0]{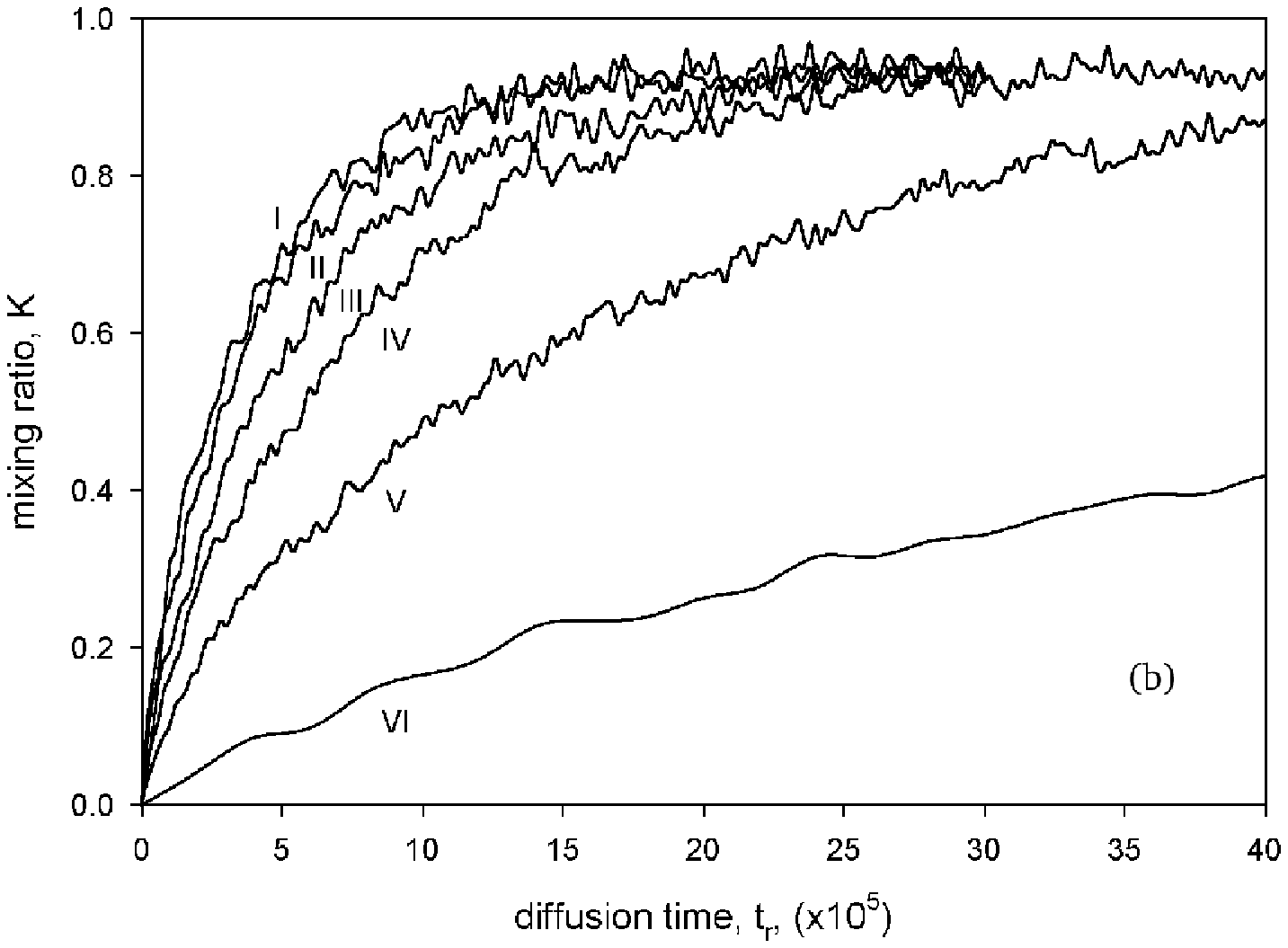}
\caption{\small{a) Variation of K with respect to
diffusion time, $t_r$ for different molecular weights, I) $M=100$,
II) $M=150$, III) $M=200$, IV) $M=250$ for $d=15$, b) The plots K versus diffusion time, $t_r$ for
different $d$ values; I) $d=25$, II) $d=20$, III) $d=15$, IV)
$d=10$, V) $d=5$ and VI) $d=1$, for $M=200$.
}}
\ec
\end{figure}
%%%%%%%%%%%%%%%%%%%%%%%%%%%%%%%%%%%%%%%%%%%%%%%%%%%%%%%%%%%%%%%%%%%%%%%%%%%%

%%%%%%%%%%%%%%%%%%%%%%%%%    FIGURE 8ab  %%%%%%%%%%%%%%%%%%%%%%%%%%%%%%%%%%%%
\begin{figure}
\bc
\leavevmode
\includegraphics[width=8cm,angle=0]{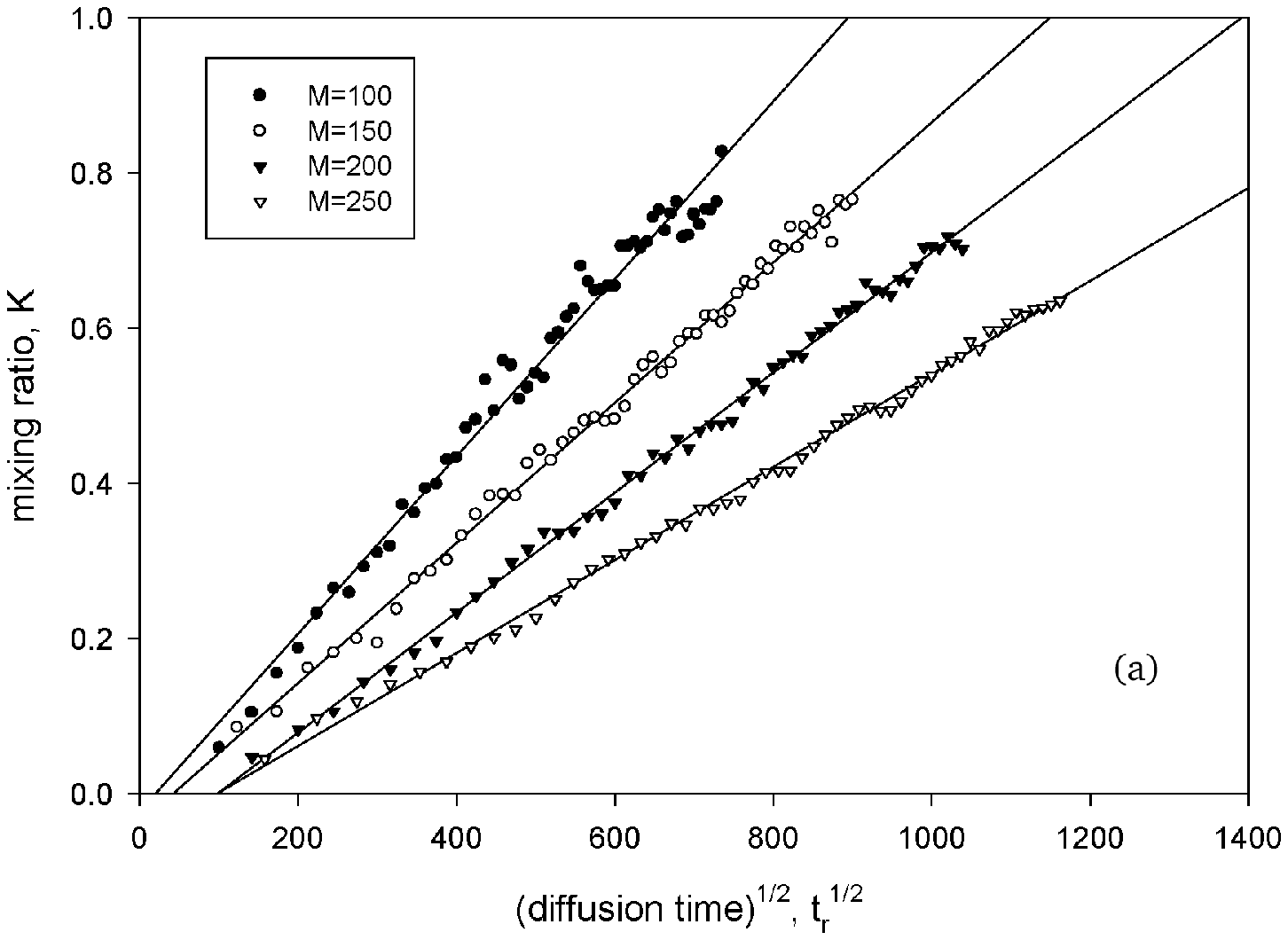}
\bigskip
\includegraphics[width=8cm,angle=0]{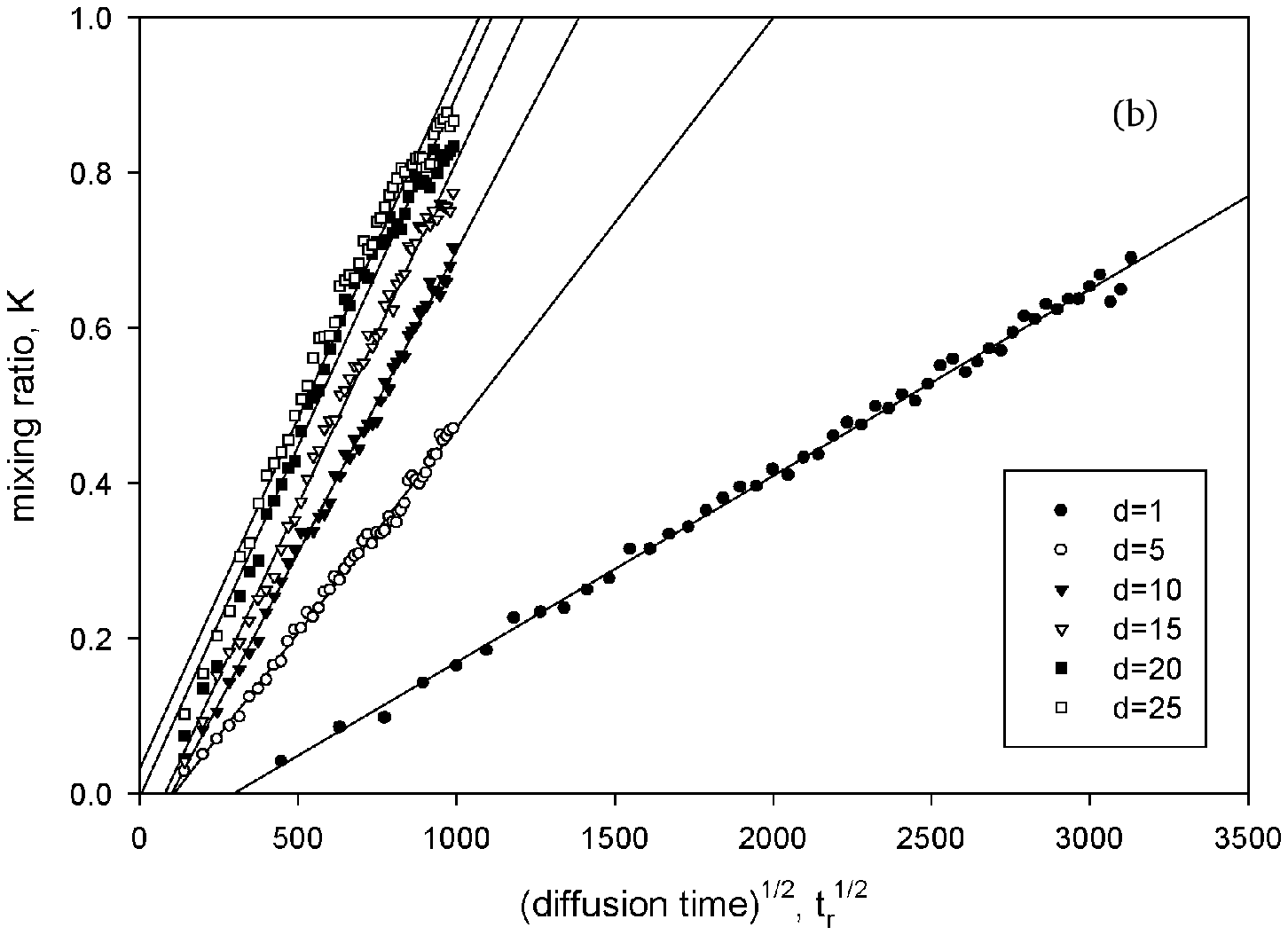}
\caption{\small{a) K versus ${t_r}^{1/2}$ plots for
different values of $M$, b) K versus ${t_r}^{1/2}$ plots for different
values of $d$. Slopes of linear fits produce $a$ in equation
(\ref{fit}).
}}
\ec
\end{figure}
%%%%%%%%%%%%%%%%%%%%%%%%%%%%%%%%%%%%%%%%%%%%%%%%%%%%%%%%%%%%%%%%%%%%%%%%%%%%

\subsection{Freely Rotated Chain Model for Reptation}

The polymer chains were constructed from the rod like segments
which were connected to each other with a bond angle, $\phi=60^o$.
The dihedral angle, $\theta$ was chosen randomly as seen in Figure
2. The length of a segment was taken to be unity and the molecular
weight, $M$ of the chain was assumed to be proportional to the
number of rods (monomers) that chain contains. Monte Carlo
simulations were performed so that the chains move according to
the reptation model, where for each reptation step, while a
segment from the tail disappears, another segment is added to head
or vice versa. In other words the chain is being trapped in a
hypothetical tube, leaves the tube in a randomly chosen
directions, creating a segment of minor chain as shown in Figure
1. A unit of diffusion time, $t_r$ was defined for a single
reptation step of a chain that has 100 monomers. For instance
single reptation step of a chain with 200 monomers takes 2 units
of $t_r$. The number of reptation steps in chosen direction for
each simulation was defined as $d$ which controls the escape time
i.e. an increase in $d$ shortens the escape time, $T_r$. Figure 3a
presents the behavior of escape time, $T_r$ against $d$ for a
polymer chain which has 200 monomer units. Here $T_r$ was
determined with a separate program which uses the same reptation
algorithm as in the simulation programs. In this program a single
chain reptates until it escapes from its initial tube and then
$T_r$ is recorded. The average $T_r$ is calculated from several
runs. It is seen in Figure 3a that as $d$ is increased $T_r$
decreases i.e. chain escapes quicker from its initial tube. The
relation between $T_r$ and molecular weight, $M$ for different $d$
values are shown in Figure 3b, where linear behavior of
$T_r^{1/3}$ versus $M$ suggest that polymer chains reptate in
accordance with equation (\ref{tr_prop_m3}) in our KMC
simulations. The steepest curve for $d=1$ predicts that
corresponding polymer chain needs longer time than others to
escape from its tube. However when chain reptates with $d=25$
escape time, $T_r$ is very short for all molecular weights.

The minor chain growth was monitored by observing the mean square
escape length $<\ell^{2}>$ versus diffusion time, $t_r$ where
$<\ell^{2}>$ was determined using the same separate program which
determines $T_r$. This program reptates a single chain and records
the minor chain length. The average was calculated after several
runs of this program. Results are shown in Figure 4a for a chain
contains 200 monomer units, where it is seen that the region
between $t_r=0$ to $t_r=0.3T_r$ presents linear relation which
accords with equation (\ref{lsquare}). The slope of the curves in
the linear region is proportional to the curvilinear one
dimensional diffusion coefficient, $D_1$ which were produced for
various molecular weights, $M$. The inverse of the slopes,
$(slope)^{-1}$ are plotted versus $M$ in Figure 4b where the
perfect linear relation is in accord with equation
(\ref{d1_prop_invM}) i.e. $D_1 \sim M^{-1}$.

\subsection{Energy Transfer for Chain Diffusion}

Diffusion of D and A labeled chains between adjacent compartments
was simulated using above KMC algorithm. The sample cube with the
side, $L=500$ units, was divided into two equal compartments which
contain donor, D and acceptor, A labeled chains respectively,
presented in Figure 5. Each compartment contains 500 chains and
one percent of monomers in each chain labeled with donors or
acceptors. Here one has to be noticed that chains are chosen quite
short and the system is very dilute for the reptation model. In
fact, here the reptation algorithm is designed so that it operates
as if the chains are in the melt system. This algorithm is chosen
to avoid the immediate decay of donors where the chains are kept
far away from each other and still perform the reptation motion.
One may attemp to solve this problem by filling up the system with
unlabeled chains by designing more realistic algorithm, which,
however then spends tremendous computer time to perform reptation
motion for all chains. Since these unlabeled chains have no
contribution to the DET, using the reptation algorithm in the
dilute system saves considerable amount of computer time and still
mimics the realistic reptation motion. The measured reptation
parameters have shown that the chosen dilute system and the
reptation algorithm work quite well and the attempt to study the
DET during diffusion is reliable.

The decay of donor intensity by DET was simulated for the possible
configurations at the end of each $10^4$ steps of diffusion,
therefore the snapshots of diffusion processes can be monitored
quite clearly and accurately. Reflected boundary conditions were
used from the sides of the cube. For the configurations of $d=1$
the decay of donor intensity was simulated after $0.5\times10^5 -
10^5$ steps of diffusion because diffusion process for this case
takes much longer time. Snapshots of the diffusing polymer chains
between adjacent compartments in a cube are shown in Figures 5a, b
and c at various diffusion steps. The D and A labeled chains are
colored in black and grey, respectively, for clarity.

The donor decay profiles were generated using equation (\ref{I2}).
The $w_{ik}$  values for each donor-acceptor pair were obtained
from equation (\ref{wik}) using a F{\"o}rster distance of 26
units. The parameter $\kappa^2$ was chosen as 0.476, a value
appropriate for immobile dyes,\textsuperscript{\cite{gunturk1}}
and the donor lifetime $\tau_0$ was taken as 44ns. Then equation
(\ref{I2}) was used to simulate the donor intensity $I(t)$ i.e.
fluorescence decay profile. $I(0)=2 \times 10^4$ was chosen and
the decay profiles were obtained for a 250ns interval, divided
into 250 channels of 1ns each. The fluorescence decay profiles
$I(t)$, for the polymer chains with $M=100$ and $d=10$ are given
in Figure 6a at various diffusion steps, i.e. various mixing
ratios, $K$. In order to obtain more realistic decay profiles,
gaussian noise was added to the original decay profiles using Box,
Muller and Marsaglia (1958) algorithm.\textsuperscript{\cite{box}}
Here, one may also take into account the effect of the lamp
profile when calculating the decay
profiles.\textsuperscript{\cite{gunturk1,gunturk2}} To do so the
decay profiles generated by the Monte Carlo simulation should be
convolved with experimental lamp profile, then the experimentally
measured $\psi(t)$ is obtained by convolution of $I(t)$ with the
instrument response function $L(t)$, as

\be \psi(t)=\int_{0}^{t} L(t)I(t-s)ds \;\; . \label{convol} \ee

\noindent In this work generated decay profiles were used, namely
$I(t)$.\textsuperscript{\cite{tuzel}} This assumption is valid if
one uses a delta, $\delta$ function light source (e.g. a very fast
laser) as the lamp profile. In this case no convolution is needed
and equation (\ref{convol}) produces $I(t)$. The noisy decay
profiles of donors for a $\delta$ function light source are
presented in Figure 6b at various diffusion steps, similar to
Figure 6a.

%%%%%%%%%%%%%%%%%%%%%%%%%    FIGURE 9ab  %%%%%%%%%%%%%%%%%%%%%%%%%%%%%%%%%%%%
\begin{figure}
\bc
\leavevmode
\includegraphics[width=8cm,angle=0]{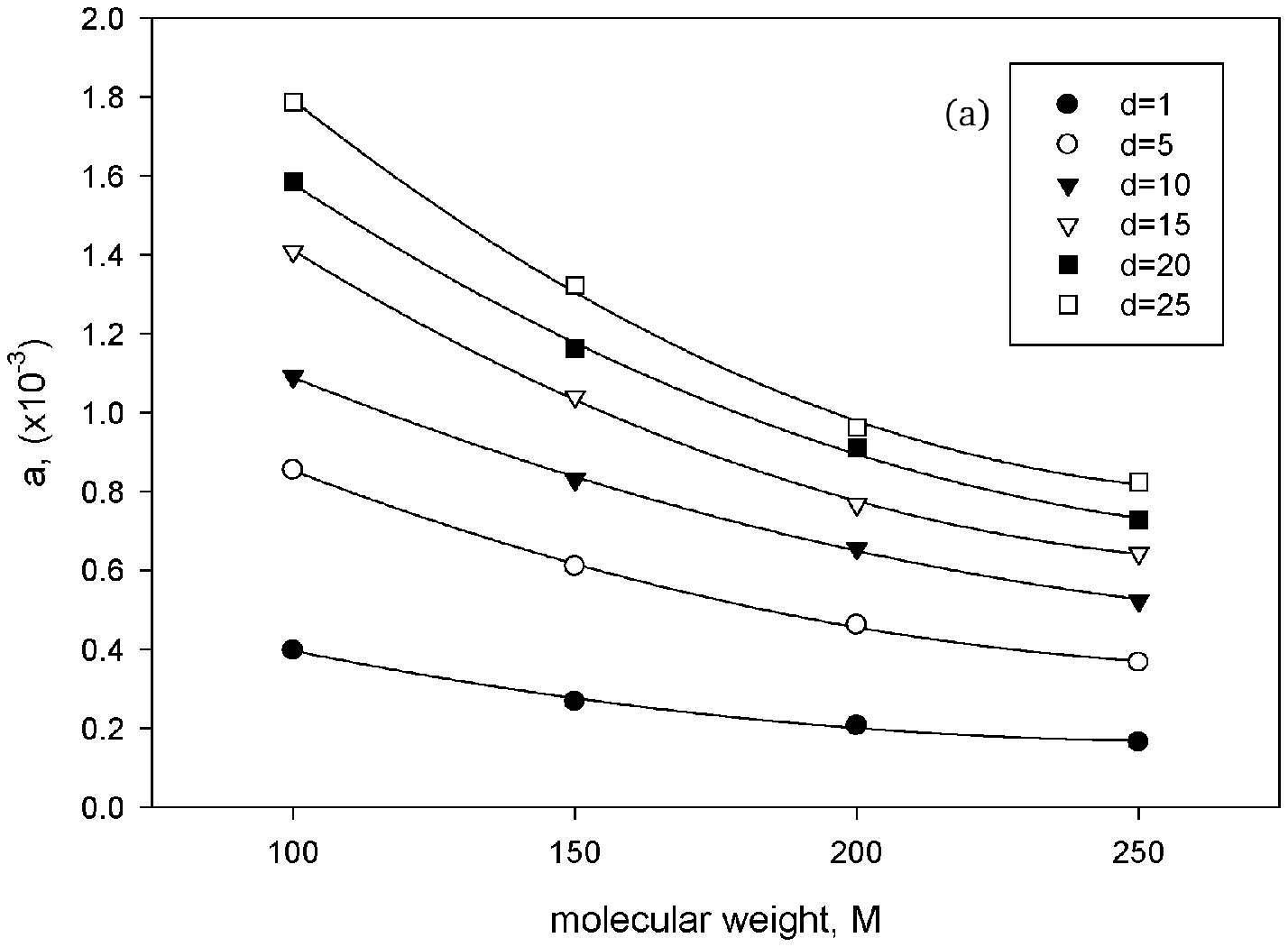}
\bigskip
\includegraphics[width=8cm,angle=0]{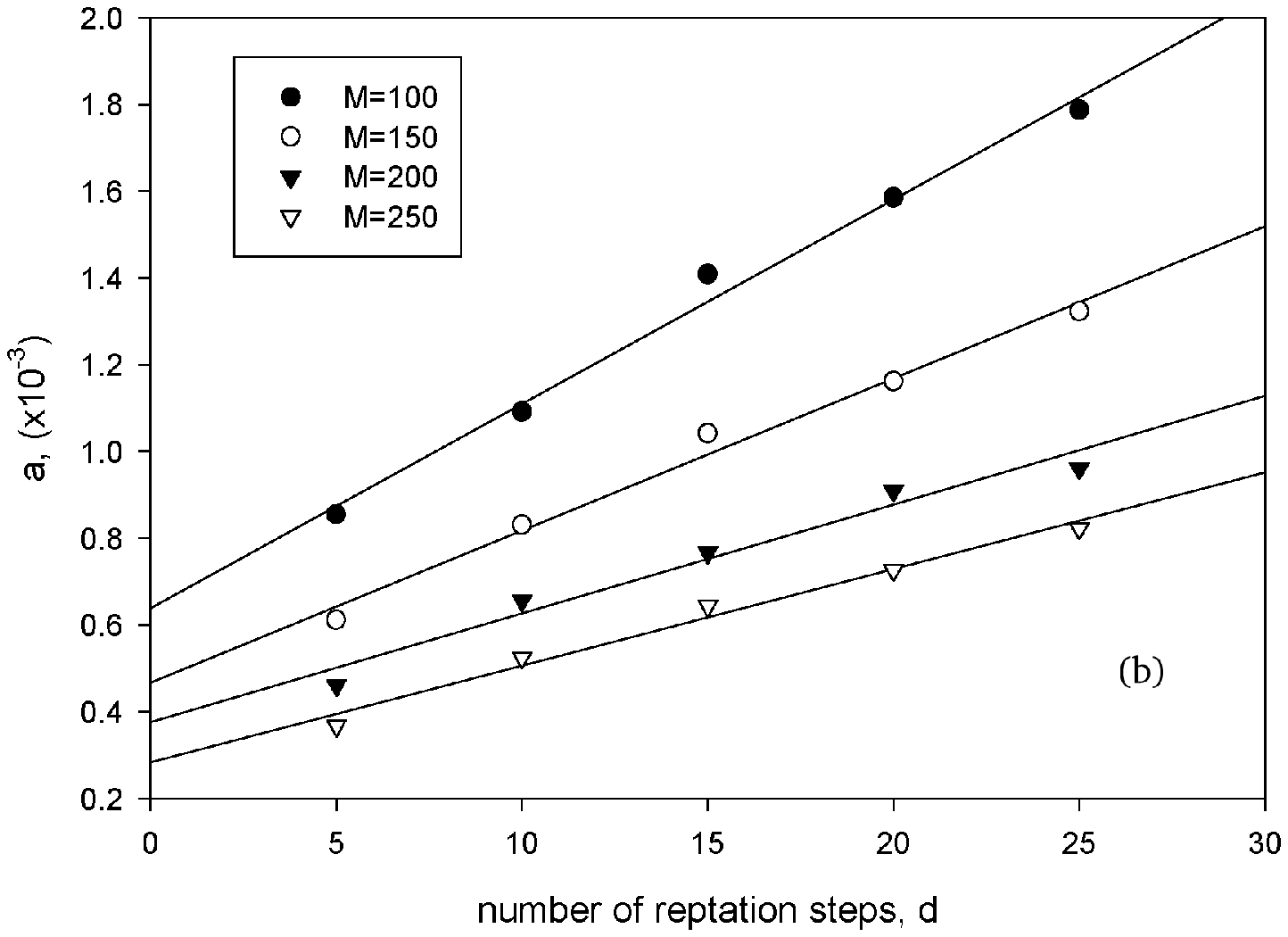}
\caption{\small{a) The dependence of $a$ on
molecular weights, $M$, for various values of $d$, b) Plots of the variation of $a$ with $d$, for
different values of $M$. The linear relation between $a$ and $d$
are shown by the regression lines.
}}
\ec
\end{figure}
%%%%%%%%%%%%%%%%%%%%%%%%%%%%%%%%%%%%%%%%%%%%%%%%%%%%%%%%%%%%%%%%%%%%%%%%%%%%

%%%%%%%%%%%%%%%%%%%%%%%%%    FIGURE 10  %%%%%%%%%%%%%%%%%%%%%%%%%%%%%%%%%%%%
\begin{figure}
\bc
\leavevmode
\includegraphics[width=8cm,angle=0]{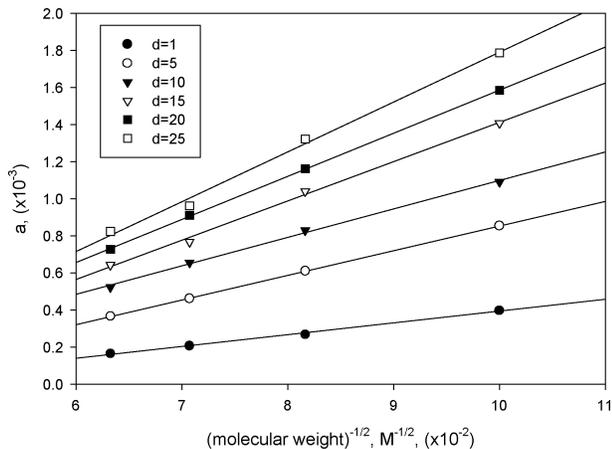}
\caption{\small{$a$ versus $M^{-1/2}$ plots for
different values of $d$. The linear regression lines show the
goodness of fit to equation (\ref{aMrelation}).
}}
\ec
\end{figure}
%%%%%%%%%%%%%%%%%%%%%%%%%%%%%%%%%%%%%%%%%%%%%%%%%%%%%%%%%%%%%%%%%%%%%%%%%%%

\subsection{Extend of Diffusion}

In order to calculate the mixing ratio, $K$ defined in equation
(\ref{k}), the decay profiles were fitted to equation
(\ref{pnemon}) using
Levenberg-Marquart\textsuperscript{\cite{bevington}} algorithm.
During curve fitting process $C$(=1) and $\tau_0$(=44) were kept
constant and $B_1$ and $B_2$ values were varied. Since $C$ and
$\tau_0$ were fixed, the fitting procedure directly produced $B_1$ and
$B_2$ values, which are a pre-exponential factor in mixed and unmixed
regions. At early times of diffusion $B_2$ dominates by presenting
low mixing, however at later times $B_1$ increases and dominates
the mixing ratio. Here basically simulation of decay curves are
essential in calculation of $B_1$ and $B_2$ values. More than
$10^4$ decay profiles were fitted and the goodness of fitting was
accepted as $\chi^2 < 1.5$. The $B_1$ and $B_2$ values
were used to obtain $K$ values at diffusion steps during reptation
of polymer chains. Figure 7a presents the variation of $K$ with
respect to the diffusion time, $t_r$ for chains at different
molecular weights, $M (100,150,200,250)$ for $d=15$. As seen in
Figure 7a mixing of diffusing chains are much faster for the low
molecular weight samples (I,II) than high molecular weight samples
(III,IV). When the molecular weight is increased diffusion slows
down as expected. Diffusion for different $d$ values is also shown
in Figure 7b, $K$ versus diffusion time are plotted for various
$d$ values $(1,5,10,15,20,25)$ for chains at $M=200$, where it is
seen that increase in $d$ cause increase in mixing of diffusing
chains. In order to quantify the above results the following
equation was employed

\be
K=a t_r^{1/2} \;\;.  \label{fit}
\ee

\noindent Here we intent to elaborate the equation
(\ref{interpenet}), where the molecular weight, $M$ can be related
with $a$ in equation (\ref{fit}). The fits of the data in Figures
7a and 7b to equation (\ref{fit}) are given in Figures 8a and 8b,
respectively. The slopes of the linear relations in Figure 8
produce $``a"$ values which are plotted versus $M$ and $d$ in
Figures 9a and 9b, respectively. Curves in Figure 9a predict that
as $M$ is increased, $a$ decreases. In order to determine the
behavior of $a$ with respect to $M$; an $a$ versus $M^{-1/2}$ plot is
presented in Figure 10 where the following relation is obeyed.

\be a \sim M^{-1/2} \label{aMrelation} \ee

\noindent Here one may speculate that mixing ratio, $K$ is most
probably proportional to the average interpenetration contour
length, $\ell(t_r)$ according to equation (\ref{interpenet}). In
other words $K$ is the measure of $\ell(t_r)$ which is quite
important in determining the mechanical properties of polymeric
materials.\textsuperscript{\cite{wool}}

On the other hand the plot in Figure 9b suggests that as $d$ is
increased, $a$ increases which predicts that as the chains reptate
faster, the mixing ratio, $K$ increases as expected. It is evident that 
since small chains reptate faster, they can mixe much
quicker than high molecular weight chains. Here one may also argue
that the root square of equation (\ref{lsquare}) suggests that
$D_1 \sim M^{-1}$ relation holds for our simulation results.

\section{Conclusion}

In conclusion this paper has presented a Kinetic Monte
Carlo method which can be used to simulate the interdiffusion
of reptating polymer chains across the interface. It
was shown that Equation (1) can be succesfully applied
to measure the extend of interdiffusion of reptating
donor and acceptor labeled polymer chains. Monte Carlo
results have shown that the curvilinear diffusion coefficient
is inversely proportional to the weight of the polymer
chains and for a given molecular weight the average
interpenetration contour length was found to be proportional
to the extent of interdiffusion.

\end{document}